# Towards a Taxonomical Consensus: Diversity and Richness Inference from Large Scale rRNA gene Analysis

Dimitris Papamichail[1,2], Celine C. Lesaulnier[1,3], Steven Skiena[2], Sean R. McCorkle[1], Bernard Ollivier[3], Daniel van der Lelie[1*]

[1]Brookhaven National Laboratory, Biology Department, Building 463, Upton, NY11973-5000, USA

[2]Computer Science Department, Stony Brook University, Stony Brook, New York, 11794, USA

[3]IRD, UMR180, IFR-BAIM, Université de Provence, ESIL, F-13288, Marseille Cedex 09, FRANCE

* Corresponding author: Daniel van der Lelie

Phone: +1 (631) 344 5349

Fax: +1 (631) 344 3407

Email: vdlelied@bnl.gov

**ABSTRACT**

Population analysis is persistently challenging but important, leading to the determination of diversity and function prediction of microbial community members. Here we detail our bioinformatics methods for analyzing population distribution and diversity in large microbial communities. This was achieved via (i) a homology based method for robust phylotype determination, equaling the classification accuracy of the Ribosomal Database Project (RDP) classifier, but providing improved associations of closely related sequences; (ii) a comparison of different clustering methods for achieving more accurate richness estimations.

Our methodology, which we developed using the RDP vetted 16S rRNA gene sequence set, was validated by testing it on a large 16S rRNA gene dataset of approximately 2300 sequences, which we obtained from a soil microbial community study. We concluded that the best approach to obtain accurate phylogenetics profile of large microbial communities, based on 16S rRNA gene sequence information, is to apply an optimized blast classifier. This approach is complemented by the grouping of closely related sequences, using complete linkage clustering, in order to calculate richness and evenness indices for the communities.

**INTRODUCTION**

Metagenome shotgun sequencing and analysis is revolutionizing the field of molecular ecology, revealing a more complete vision of the biodiversity and functions within ecosystems (2-6, 12, 29-31). Preliminary to constructing and sequencing metagenome libraries, it is necessary to estimate the complexity and richness of the microbial community to be examined. This can be adequately done by constructing and sequencing ribosomal RNA (rRNA) gene libraries (8). Thus, the means to accurately identify and classify large datasets of rRNA genes and to determine community richness is an essential step in metagenome sequencing projects. The increasing library sizes to cover the metagenomes of complex microbial communities and the subsequent amount of screening and analysis needed necessitates the development and application of adequate and robust bioinformatics tools for community composition analysis that can handle large rRNA gene data sets.

The following tools are available for analyzing microbial community composition:

**Phylogenetic trees** are useful in determining novel groups of organisms, but are limited in their reliability for large datasets. Bootstrap values become expensive to calculate and tree topologies become unreliable as tree size increases (20).

**Oligonucleotide (k-mer) based classifiers** provide fast and reliable techniques, which exploit the fact that closely related sequences share common small subsequences and that organisms and regions have their own distinct signatures, a term to describe distributions of oligonucleotides (10, 14, 22, 23). A disadvantage is the lack of positional information of oligonucleotides, which becomes more problematic with decreasing k, in which case the probability of finding oligos by chance increases exponentially. Adding location information to oligonucleotides would lead to computation time increases, while limiting its generality. In addition, altering a single base in a sequence results in $k$ different $k$-mers, and random mutations can therefore have devastating results on the oligonucleotide distribution. Correcting this problem by approximate matching for $k$-mers leads to exponential time increases relative to the number of errors allowed.

At present, the most widely used tool for species identification based on ribosomal RNA gene sequences is the ribosomal database project (RDP) Naïve Bayesian classifier (9). This tool uses 8-mers to cluster and classify 16S rRNA sequences based upon vetted sequences with well assigned taxonomy (http://rdp.cme.msu.edu/index.jsp, James Cole, private communication) (11).

**Sequence alignment classifiers** extract differences and calculate distances between DNA sequences. They can be used to identify the closest match of a sequence to a vetted reference data set. The computation time required for aligning sequences increases quadratically to the length of the sequences, which significantly delays the analysis of large datasets, e.g. rRNA gene sequences representing a complex microbial community.

The program BLAST (1) performs approximate sequence alignments, finding locally very similar pieces as opposed to globally calculating the best way to convert one sequence to the other. BLAST can be calibrated to achieve a desired speed/sensitivity ratio. Local alignment provides flexibility in handling sequencing errors, incorrectly inserted or omitted prefixes, suffixes and subsequences, as well as ambiguous characters. A further advantage of BLAST is the ability to compare sequences, one at a time, against a database of vetted sequences. Classification using alignment is highly parallelizable, with great promise over multi-processor and future multi-core systems.

90

91   Two important factors that describe a microbial community are *richness*, meaning the
92   number of species present, and *diversity*, which is their relative abundance (21). The
93   latter can be estimated from the classification efforts and/or phylogenetic analysis of
94   community samples. Richness estimation requires information on the number of
95   distinct subpopulations present in the community, according to a threshold set to
96   determine them (e.g. genus level), as well as the evenness information, meaning how
97   different the sizes of the subpopulations in the community are. Several richness
98   estimator methodologies have been developed including extrapolation from
99   accumulation curves, parametric estimators and non-parametric estimators, the latter
100  being the most promising for microbial studies (13). Among this last class of
101  estimators, Chao1 (7) seems to be the most suited method for estimating phylotype
102  richness from prokaryotic 16S rRNA libraries (16).

103

104  In this manuscript we describe the use of local alignment classification as a more
105  robust tool in grouping closely related sequences, while maintaining equivalent or
106  slightly increased classification accuracy as compared to the RDP naïve Bayesian
107  classifier. In addition, we argue against the single linkage clustering methodology for
108  richness estimation calculations in favor of the equally computationally-attractive
109  complete linkage method.

110

## MATERIALS AND METHODS

**Computation**

All computational experiments were performed on either a hyper threaded Pentium 4 at 3.2GHz desktop with 2GB of memory or dual Xeon at 2.8GHz server with 4GB of memory. Since most classification computations are parallelizable, a cluster of 4-10 desktops was used for reducing computational time. All time references referring to computation will assume use of one CPU desktop with at least 2GB of memory.

All programs/scripts performing computations and statistical evaluations were written in Perl, except the edit distance calculations, which were written in C in order to decrease computational time. The program R (28) was used for the clustering analysis. Computation times for the single, average and complete methods differ insignificantly and have a quadratic time dependence to the number of sequences (or linear to the number of pair-wise sequence distances). Evaluation of the average and maximum in-group distances of a clustering were performed using the height and order arrays provided by R, in conjunction with the calculated pair-wise edit distances of the vetted sequences.

**16S rRNA Sequence Classification**

Classification accuracy was measured by performing leave-one-out tests of the 5574 vetted sequences against themselves. A subset of vetted sequences is created, excluding the singletons, which are sequences that belong to a phylogenetic group with only one member. Each sequence of this new dataset is then separated from the dataset and classified against it. The number of correctly classified sequences is then divided by the total number of sequences present in this subset (5246 - with singletons being excluded) to produce the final accuracy percentages.

Classification with the RDP naïve Bayesian classifier takes approximately 1.5 minutes, when submitted through the web, to produce a complete 1000 sequence assignment with confidence estimates, as calculated by bootstrap trials.

The *blastn* classifier requires, for the same number of sequences and against the same database of vetted sequences, approximately 35 minutes.

Using *bl2seq*, for performing pair wise comparisons, requires significantly more time, in the range of days.

Increasing the accuracy of BLAST by lowering the values of the word size $W$ and the $X$ *drop-off* parameters results in a significant increase in computation, while increasing the classification accuracy at the genus level by no more than 0.1%.

The naïve Bayesian classifier principles are described in (23). The RDP classifier uses oligonucleotides of size 8 and randomizes the selection of oligonucleotides to be used for the confidence calculation. In each of the 100 bootstrap trials, 1/8 of all possible 8-mers of a sequence are selected randomly.

The BLAST parameters adjusted for *bl2seq* and *blastn* in order to minimize misclassifications were:

1. *match*: The reward for a matched character, common to both sequences compared.
2. *mismatch*: The penalty for a character substitution
3. *gap_start*: The penalty for initiating a gap.

161    *gap_extend*: The penalty for extending an initiated gap by one character.

162    These values can be scaled proportionately without affecting the alignment, but only

163    the score, although the relative scores under the same parameter set remain

164    proportional.

165

166    The score for each BLAST alignment, used to determine confidence values, was

167    calculated by summing up the individual scores of the locally aligned pieces, which is

168    already normalized against the length of the sequences being compared. The two tests

169    to measure misclassifications and calibrate local alignment parameter space were (i)

170    the number of rRNA sequences that score better against a sequence from another

171    genus than against all of the sequences in their genus and (ii) the number of ribosomal

172    sequences that score better against a sequence from another genus than at least one

173    sequence from their own genus. The first test was used predominantly, since the final

174    classification decision is based on the top scoring vetted sequence and can lead to a

175    misclassification only if the test sequence aligns with a higher score against a foreign-

176    genus vetted sequence.

177    For scoring an alignment, we used the sum of scores of individual local aligned

178    pieces, which are not overlapping. A query sequence is assigned the phylogenetic

179    lineage of the highest matching score.

180

181    **Levenshtein edit distances**

182    The Levenshtein edit distance (17) between two sequences is defined as the number

183    of edit operations – insertions, deletions and substitutions – required to transform one

184    sequence to the other, where each edit operation has a cost of 1.

185

186 **RESULTS**

187 **RDP naïve Bayesian classification and initial drawbacks**

188 We initially tried to use the Ribosomal Database Project (RDP) Naïve Bayesian
189 classifier to analyze the 2774 16S rRNA gene sequence data from a large scale
190 sequencing project, which aimed at determining changes in the soil microbial
191 community composition of trembling aspen when three were exposed to ambient (360
192 ppm) or elevated (560 ppm) atmospheric $CO_2$. Upon initially analyzing 2774 16S
193 rRNA gene sequences it became apparent that a large number of them, which were
194 classified in the same genus, had edit distances accounting for >20% of the total
195 sequence length, and sometimes up to 50%. Many of these occurred between
196 sequences classified with low confidence estimates (<50%), creating uncertainty for a
197 large number of classification groupings. This provided the first clue that many of the
198 sequences we were trying to classify were distant from all the vetted sequences
199 available, possibly representing new phylogenetic groups. We also noticed that some
200 sequences with >99% edit distance similarity were found classified in different taxa,
201 with a couple of occurrences even at the phylum level. These findings bring to light
202 the fact that a perfect classification of rRNA gene sequences can currently not be
203 achieved, and that errors will be found even when classifying unknown sequences to
204 closely related characterized ones. These results prompted us to explore sequence
205 identification methods alternative to the RDP Naïve Bayesian classifier.

206

207 **Refinement of Vetted Sequence Classification using BLAST**

208 The BLAST utility bl2seq (27), which performs pair wise sequence alignment, was
209 used as an alternative to maximize 16S rRNA gene classification accuracy. Several
210 key parameters (match, mismatch, gap_start, gap_extend) were adjusted to minimize misclassifications. To do so, a genus level leave-one-out test for each of the 5574 vetted sequences (RDP data set) was performed to determine the parameter set that optimally separates the scores of closely related species from distant ones. After examining the different parameter sets we determined the optimal set of key parameters (*match* = 1, *mismatch* = 5, *gap_open* = 3, *and gap_extend* = 2.5), where the *match* is positive and considered a reward, and *mismatch*, *gap_open* and *gap_extend* are negative and are considered penalties. Compared to the default bl2seq parameter values, this set of parameters reduced the number of misclassified vetted sequences from 284 to 268 out of a total of 5246 (328 sequences are unique in their genus in the vetted set).

**Classifier confidence levels**

We created a confidence estimate for each bl2seq alignment score using the value of the highest scoring pair-wise alignment for each sequence, setting the Boolean value for correct (1) or incorrect (0) classification at the genus, family, order, class or phylum level. This gives us the ability to assign a confidence estimate to each specific score value, according to the number of times an alignment with such a value resulted in the correct phylogenetic classification. A rating for this score was then based on these results. Fourth degree polynomial regression curves were used to determine the relation between classification scores and confidence estimate values for the different phylogenetic levels. These curves (Figures 1a-e) were, smoothed by the addition of extra points, representing high confidence estimates at very high scores and zero confidence estimates at very low scores. The confidence estimate of each score is calculated from the value of the polynomial for this specific score. From this analysis we can conclude that confidence estimates decrease when classifying the species at lower phylogenetic levels (from phylum to genus). Figure 1f demonstrates that we still obtain a 94% classification accuracy at the genus level, when using the optimal parameter set and the bl2seq utility, and increasing accuracies for higher phylogenetic levels.

**Exploring the alignment parameter set**

We subsequently constructed a vetted sequences database in BLAST format and used the *blastn* utility, as previously done for bl2seq, in a leave-one-out test on this database. The faster *blastn* processing of sequences, compared to bl2seq, was used to explore all possible combinations of key parameters. In order to identify the best set of parameters, we performed a full coverage scan at value increments of 5, fixing the reward value of match to 10, and allowing for all possible combinations of the three other parameters (*mismatch, gap_open and gap_extend*). Restrictions in the *blastn* utility did not allow ratios of mismatch/match lower than 1 (except the ratio 8/10) and higher than 5. Also the mismatch/match ratio of 9/2 was not permitted. We observed that when using *blastn*, the values for *gap_start* and *gap_extend* did not alter our results, this in contrast to their influence on scores using the *bl2seq* utility. At the end, the (*match = 10, mismatch = 50*) assignment gave the best classification accuracy, with percentages for the different phylogenetic levels shown in Table 1, together with the published accuracy results of the RDP naïve Bayesian classifier (32).

**Improved grouping of closely related sequences using the *blastn* classifier**

Although the accuracy results presented in Table 1 do not differ significantly for the different methods tested, our *blastn* classifier groups closely related sequences with increased accuracy. To demonstrate this, Levenshtein pair wise edit distances (17) were calculated for a set of 2774 16S rRNA gene sequences representing the microbial communities associated with trembling aspen under conditions of ambient and elevated CO2 (C. Lesaulnier, D. Papamichail, S. McCorkle, B. Ollivier, S. Skiena, S. Taghavi, D. Zak, D. van der Lelie. (2007). Elevated $CO_2$ Affects Soil Microbial Diversity Associated with Trembling Aspen. Environ. Microbiol., under review). Using the complete linkage method these were subsequently clustered into groups, in which the percentage difference among sequences is below a cut-off value. Taking as an example a 1% cut-off sequence difference (this percentage is calculated proportional to the sequence length, which for the 16S bacterial rRNA gene is approximately 15), we would expect all group members to belong to the same phylogenetic group, ranging from genus to phylum, since at the 1% dissimilarity level even species are expected to cluster together. Considering the high identification percentages for both the RDP classifier and the *blastn* classifier, we counted the number of groups that were heterologous at a given phylogenetic level (e.g. contained members of more than one phylotype). A few indicative results are the following: (i) For the ambient $CO_2$ community 16S rRNA sequence set, which comprised of 132 groups with more than one element, 23 groups had elements classified in different genera by the RDP classifier, breaking them into 52 subgroups. The blastn classifier divided only 15 groups into 30 subgroups. In 14 out of the 15 groups this was due to the presence of a single misclassified sequence. The biggest group of 32 elements was identified as 32 *Desulfotomaculum* by our classifier, where the RDP broke it into 6 distinct groups, partitioned as: 4 *Thermodesulfovibrio*, 1 *Succiniclasticum*, 15 *Gelria*, 1 *Propionispora*, 3 *Thermovenabulum* and 8 *Pelotomaculum*. (ii) For the elevated $CO_2$ community 16S rRNA sequence set, 28 groups out of a total of 87 groups were divided into 67 subgroups by the RDP classifier, where the blastn classifier only divided 10 groups into 20 subgroups. These examples show that the blastn classifier reduces classification ambiguities compared to the RDP classifier. The complete results for all phylogenetic levels and with dissimilarity percentages ranging from 1%-5% can be found in the supplemental online material.

**Clustering methodology for richness estimation**

To further deconvolute community composition it is necessary to calculate the richness of a microbial population, meaning the number of phylotypes present. This requires partitioning the sampled sequences into sets according to their similarity. This can theoretically be achieved by using the output of our classifier, where information is known for the identification of all sequences at different phylogenetic levels. However, this would require that all sequences are identified with the same confidence level. This is not the case. In addition, highly dissimilar sequences can sometimes be classified in the same phylogenetic group when they have the same vetted sequence as their closest neighbor.

We examined three traditionally used clustering methods, the single linkage, average linkage and complete linkage methods, all which fall under the agglomerative hierarchical (bottom-up) approach and produce clustering trees (15). We compared the three methods using the set of 5574 vetted sequences, for which phylogenetic information for all phylotypes is well assigned. Before applying the clustering methods, we used the known phylogenetic partitioning of the vetted sequences to calculate statistics about the number of groups they form at each phylogenetic level, as well as the minimum, average and maximum Levenshtein edit distances (17) between sequences in these phylogenetic groups. The means of all these values are shown in Table 2. All vetted sequence statistics, including details about all groups, can be found in the supplemental on-line materials.

According to the number of groups in each phylotype (see Table 2), we determined the necessary cut off edit distance value, which, when applied to the inferred clustering tree, would produce the same number of phylogenetic groups. This is demonstrated in Figure 2 on a random subset of 100 vetted 16S rRNA sequences. In this figure, for example, we can observe that a cut-off edit distance value of 300 will result in the formation of 23 groups for the given 100 sequences. Inversely, if we want to acquire 15 groups, a cutoff edit distance of 380 is required. Knowing the number of distinct groups for all taxa for our vetted sequence set allows us to determine cutoff levels that will generate the same number of groups, when clustering these sequences with our three hierarchical clustering methods. This allows the evaluation of the clustering methods independently of the error in the cut-off estimation, which is actually a separate problem for all clustering and partitioning methods, and usually is calculated based on observations (13).

Correct cutoff values, as shown in Figure 2, cannot be calculated directly from vetted sequence statistics. To illustrate this point, one would expect that for the complete linkage clustering method, the correct threshold could be determined by calculating the maximum in-group distance, when the number of groups formed is the same as in the vetted sequence set at some phylogenetic level. It happens though that, even at the genus level, a group exists (Clostridia) in the vetted sequence set with a maximum in-group distance of 626, which indicates approximately 43% sequence dissimilarity.

More appropriate thresholds can be determined by considering the mean of the average distances inside the groups at every merging step of the average linkage clustering hierarchical algorithm, and then comparing this value to the known mean for the corresponding groups of the vetted sequences. These values are quite similar, as can be seen in Table 3. The same effect is observed for the complete linkage clustering method (Table 3), where the threshold value for partitioning is determined based on the maximum distance inside each group. Single linkage clustering does not offer such a measure for estimating a cut-off, since there is no averaging process in the algorithm.

By sorting the groups, formed at the different phylogenetic levels by using the different clustering methods, according to their cardinality, and by comparing this to the known phylogeny, we created graphs showing the trends in group sizes. The known phylogeny group cardinalities were the best approximated by the complete, followed by the average clustering method. This is shown in Figure 3 for the 75 groups at the order level. Similar figures for all taxonomic levels can be found in the supplemental on-line materials.

To quantify the better performance of the complete clustering method, as observed in Figure 3, we calculated the Pearson correlation (difference in variance) and square difference (distance of each individual group of the same index, according to the corresponding sorted position). The results are presented in Table 4 and confirm that the complete linkage clustering method provides a better correlation to the known classification at all five phylogenetic levels. As a case study for the richness estimation based on different clustering methods, we calculated the Chao1 index at different phylogenetic levels. Since the Chao1 estimate is based on the ratio of singletons and doubletons of a sequence grouping, it can vary significantly with changes to these small integers. For that reason, we tested the Chao1 richness estimations on 1000 random selected sequences from the vetted sequence set, repeating the test a total of 1000 times. In Table 5 we present the average richness

360    estimations of these experiments. As seen in Table 5, richness estimations based on
361    groups clustered with the complete linkage method are the most accurate. In
362    conjunction with the consistently better correlation of the group size histograms,
363    complete linkage clustering is preferable for use in richness estimation analysis.
364
365

**DISCUSSION**

The 16S rRNA gene sequence enables the association of phylogeny (18, 26) and remains the most reliable method to determine completely new or divergent organisms. Aside from the availability of a curated dataset (i.e. the vetted sequences), the analysis of 16S rRNA gene sequences serves as a choice model, as it also permits a direct comparison of the composition of different communities.

Although recent tree manipulation and visualization utilities like *arb* (19), which use multiple alignment to construct phylogenetic trees, have the capability of handling large datasets, editing their input becomes a laborious and tedious task. Therefore, the need exists to develop classification tools to overcome both the computational limitations in accurately identifying taxonomical relationships, and reconstructing phylogenetic trees for the purpose of better extrapolating ecological roles. We developed a *blastn* classifier with optimal key parameter set that performs better than the RDP II classifier for 16S rRNA based identification, especially when it comes to grouping of related sequences, thus reducing classification ambiguities. However, every classifier has a closed architecture and will assign every sequence to one in its dataset. The view of the biodiversity contained within a sample is therefore subject to the biases incurred by the limited number of sequences contained within the vetted sequence database, against which we classify.

Because of its simplicity and efficiency, single linkage clustering has often been used for clustering sequences (25). Other tools, such as DOTUR (24), give the user the option to select different clustering methods, but no information is provided on which method actually performs better or what dissimilarity cutoff should be used to differentiate groups at a given phylogenetic level. We demonstrate that the complete linkage clustering method seems to be the preferential approach to create clusters of closely related sequences, taking into account that it is less computational intense than full phylogenetic tree analysis. The output of this clustering method can subsequently be used for richness estimation of the microbial community, using e.g. the Chao1 index, as we did for the different microbial communities associated with trembling aspen under conditions of ambient and elevated $CO_2$.

In conclusion, our *blastn* classifier with optimal key parameter set has been proven to provide consistent and robust analysis. Further improvements could be realized in both accuracy and speed, especially through the contributions of advances in parallel and core architectures. These developments should enhance significantly the utility of database search and taxonomic annotation methods to the molecular biologist.

## ACKNOWLEDGMENTS

This research was funded by the US Department of Energy's Office of Science (BER) project number DE-AC02-98CH10886, entitled "Composition of Microbial Communities for In Situ Radionuclide Immobilization Projects", and by Laboratory Directed Research and Development funds at the Brookhaven National Laboratory under contract with the U.S. Department of Energy. We would like to thank Dr James Cole from RDP for providing us with the vetted sequence data.

513  **Table 1:** Classification accuracy of RDP, blastn and bl2seq classifiers for different
514  phylogenetic levels.

| Method | Phylum | Class | Order | Family | Genus |
| --- | --- | --- | --- | --- | --- |
| RDP naïve Bayesian classifier | 99.9 | 99.9 | 99.3 | 97.1 | 94.3 |
| blastn classifier | 99.9 | 99.7 | 99.4 | 97.1 | 94.9 |
| bl2seq classifier | 99.9 | 99.7 | 99.4 | 96.9 | 94.7 |

515  The numbers represent percentages of sequences correctly classified in their known
516  phylogenetic levels in leave-one-out tests.

517

518

519  **Table 2:** Phylogenetic partitioning of the vetted sequences in groups and their
520  statistics.

| Level | Total number of groups | Mean minimum in group distance | Mean maximum in group distance | Mean average in group distance |
| --- | --- | --- | --- | --- |
| Phylum | 30 | 36 | 399 | 233 |
| Class | 39 | 26 | 310 | 234 |
| Order | 75 | 15 | 336 | 187 |
| Family | 192 | 25 | 265 | 153 |
| Genus | 769 | 39 | 143 | 93 |

521  The known phylogenetic partitioning of 5574 vetted sequences was used to calculate statistics
522  about the number of groups they form at each phylogenetic level, as well as the minimum,
523  average and maximum Levenshtein edit distances between sequences in these phylogenetic
524  groups.

**Table 3:** Estimated cut-off values based on the vetted sequence statistics and actual cut-off values for different phylogenetic levels determined using the average and complete linkage clustering methods.

| Cut-offs / Phylogenetic level | Phylum | Class | Order | Family | Genus |
|---|---|---|---|---|---|
| Average linkage clustering mean value: Estimated/Actual | 233/231 | 234/223 | 187/208 | 153/158 | 93/95 |
| Approximate sequence dissimilarity cutoff for average linkage clustering | 15.9 % | 15.4 % | 14.3 % | 10.9 % | 6.6 % |
| Complete linkage clustering mean value: Estimated/Actual | 399/414 | 310/377 | 336/339 | 265/258 | 143/139 |
| Approximate sequence dissimilarity cutoff for complete linkage clustering | 28.6 % | 26.0 % | 23.4 % | 17.8% | 9.6 % |

The estimated cut-off values are the means of the vetted sequence statistics for in-group average and maximum sequence distances (See Table 2). Sequence dissimilarity cutoffs are presented as edit distance over average 16S sequence length percentages.

532 **Table 4:** Pearson correlation and square differences of the sorted cardinality lists of
533 partition groups, created by the three clustering methods, against the original
534 partitions of the vetted sequences

|        | Correlation |         |          | Square distance |         |          |
|--------|-------------|---------|----------|-----------------|---------|----------|
|        | Single      | Average | Complete | Single          | Average | Complete |
| Phylum | 0.8018      | 0.8041  | 0.9423   | 3159            | 2877    | 1148     |
| Class  | 0.6431      | 0.8689  | 0.9809   | 3973            | 1813    | 364      |
| Order  | 0.7939      | 0.9413  | 0.9809   | 3612            | 937     | 323      |
| Family | 0.7841      | 0.8497  | 0.9606   | 1618            | 1087    | 284      |
| Genus  | 0.8085      | 0.9769  | 0.9883   | 816             | 169     | 109      |

535 The Pearson correlation and square differences were calculated for different
536 phylogenetic levels.

537 **Table 5:** Average Chao1 richness estimation index calculated for random 1000
538 sequence subsets from the RDP vetted sequence dataset and for the groupings from
539 different clustering methods.

|  | Average of existing groups | Chao1 index on actual data | Deviation of Clustering Methods richness estimation | | |
| --- | --- | --- | --- | --- | --- |
|  |  |  | Single linkage | Average linkage | Complete linkage |
| Phylum | 21.7 | 25.1 | 38.7 % | 13.7 % | 11.8 % |
| Class | 29.8 | 33.4 | 44.3 % | 11.5 % | 8.9 % |
| Order | 63.5 | 56.6 | 51.5 % | 19.0 % | 7.3 % |
| Family | 192 | 247.9 | 63.8 % | 26.2 % | 6.7 % |
| Genus | 769 | 1281.3 | 36.3 % | 9.36 % | 11.1 % |

540

541 The first two columns present the average number of phylotypes in the 1000
542 randomly selected sequences and the estimated Chao1 richness for each phylogenetic
543 level, based on the known taxonomical grouping. The last three columns present the
544 average deviation of the Chao1 richness estimate, based on the groupings acquired
545 from different clustering methods, as a percentage of the Chao1 richness estimate of
546 the known taxonomy. Here complete linkage clustering is outperforming the other
547 clustering methods in all but one level (genus).

**Figure legends**

**Figure 1(a-e):** Fourth degree smoothed polynomial regression curves and classification score confidence estimates for different phylogenetic levels (phylum, class, order, family, genus). The confidence estimate of each score is calculated from the value of the polynomial for this specific score.

**Figure 1(f)**: Classification accuracy at different phylogenetic levels when using the optimal parameter set (*match = 1, mismatch = 5, gap_open = 3, and gap_extend = 2.5*) and the BLAST bl2seq utility.

**Figure 2:** Complete linkage clustering of a subset of 100 vetted 16S rRNA sequences (for demonstration purposes). For two different cut off edit distance values of 300 and 380, the set is partitioned into 23 and 15 groups, respectively.

**Figure 3:** Cardinalities of the 75 groups formed by the single, average and complete clustering methods, compared to the original 75-group partitioning of all 16S vetted sequences, at the order level. The logarithmic values of the group sizes are presented in reverse sorted order. The two rightmost "steps" of each curve show the number of the doubletons and singletons, representing groups with two and one members, respectively.

569  Figure 1
570

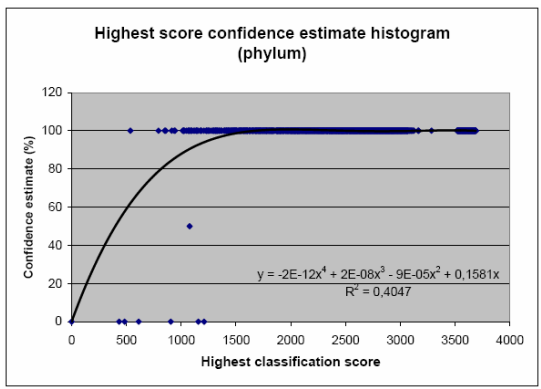

(a)

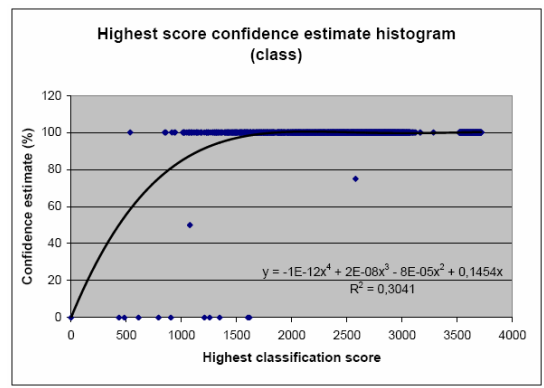

(b)

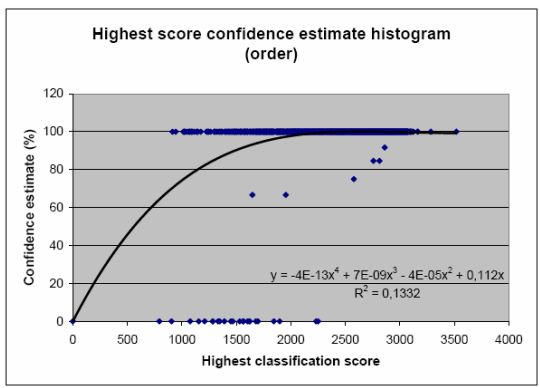

(c)

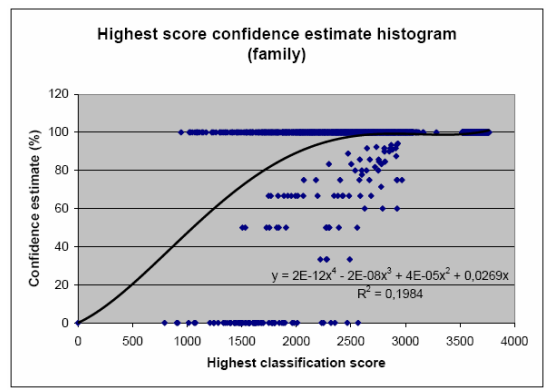

(d)

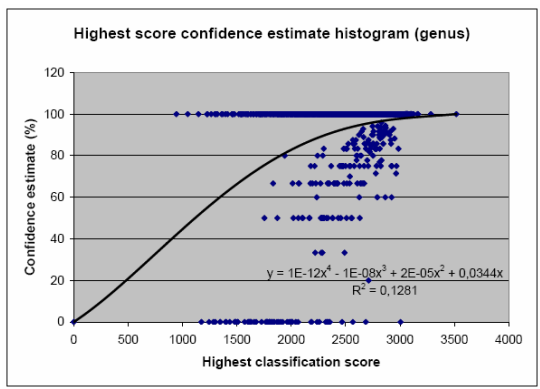

(e)

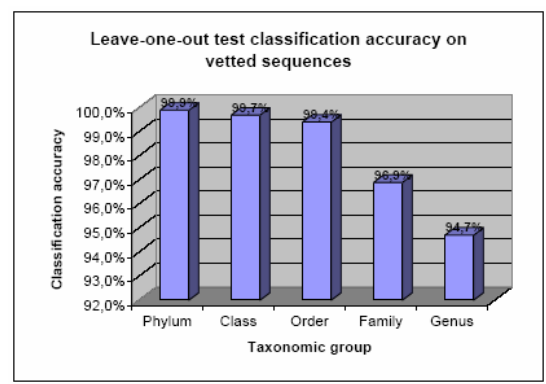

(f)

571  Figure 2:
572

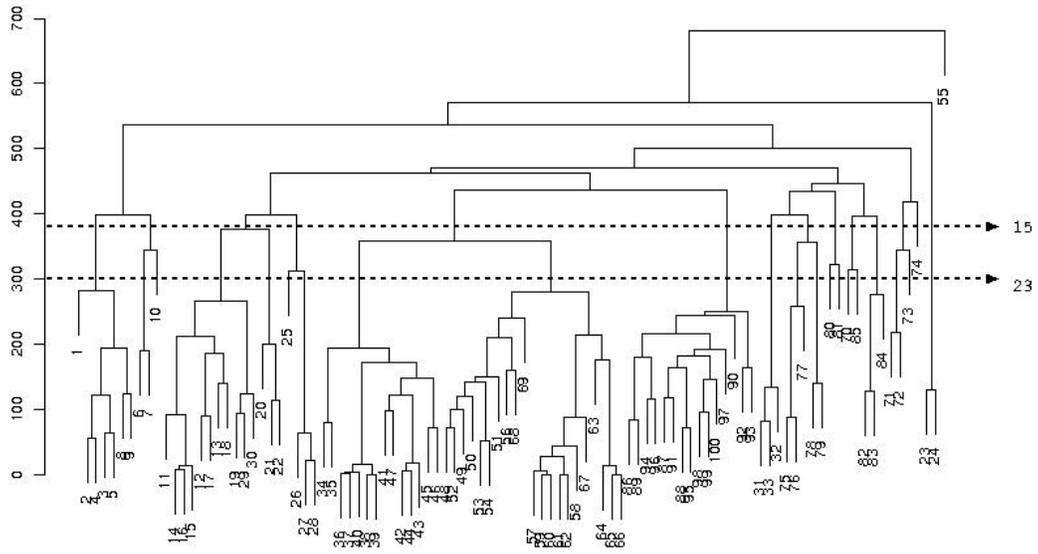

573

574     Figure 3:

575

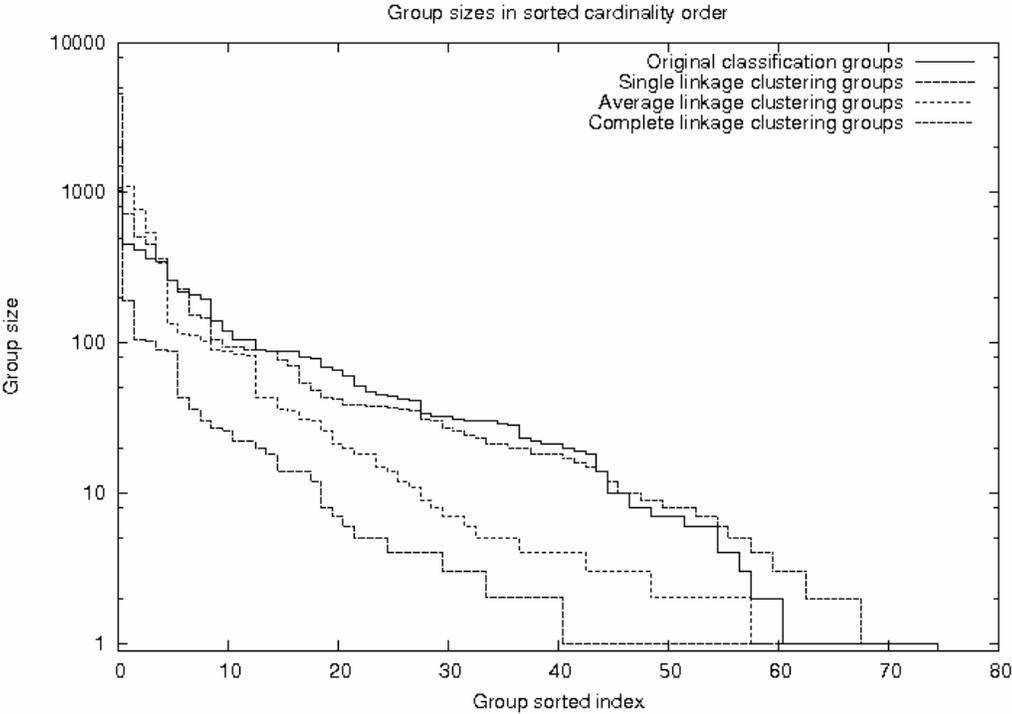

576
577
578